Failure in complex social networks

Damon Centola

*Institute for Quantitative Social Science, 1730 Cambridge St., S408, Harvard University, Cambridge, MA 02138.*

Tolerance against failures and errors is an important feature of many complex networked systems [1,2]. It has been shown that a class of inhomogeneously wired networks called "scale-free"[1,3] networks can be surprisingly robust to failures, suggesting that socially self-organized systems such as the World-Wide Web, the Internet, and other kinds of social networks [4] may have significant tolerance against failures by virtue of their scale-free degree distribution. I show that this finding only holds on the assumption that the diffusion process supported by the network is a simple one, requiring only a single contact in order for transmission to be successful.

For complex contagions [5], such as the spread of cultural norms, collective behaviour or cooperation, multiple sources of reinforcement are needed for transmission to be successful [5, 6]. On networks with high levels of local clustering, as is typical of social networks [7-9], a scale-free degree distribution makes the social topology much more sensitive to failure due to accidents and errors than does a more homogeneous, exponential degree distribution.

Exponential and scale-free networks are compared by using networks of the same size ($N$=10,000), average degree ($<k>$=4), and level of clustering ($C$=.25)[6]. Error tolerance is tested by randomly removing a fraction, $f$, of nodes from the network [1], and then measuring the average size of cascades, $S$ (the number of nodes reached by a contagion), which originate from a randomly chosen seed neighbourhood. This process is repeated over 1000 realizations to produce an ensemble average cascade size, $<S>$, for each value of $f$.

In order for complex contagions to propagate across a social network, the network not only must remain connected, but it also must have sufficient local structure [5] (Fig. 1a). When a scale-free network suffers random errors, it quickly reaches a critical fraction of removed nodes, $f_c \sim .0002$, above which cascades can only reach less than half of the network (Fig. 1b). This is independent of whether nodes are removed by targeted attack (removing the most connected nodes first) or random failure. This weakness is endemic to clustered scale-free networks because of the large fraction of the population with minimal connectivity.

The exponential network has more nodes with moderate degree, and thus there are many redundant pathways for local reinforcement. The formation of "bottlenecks" (illustrated in Fig. 1a) limits complex contagions to reaching only 70% of the entire network. Despite this, cascades can still reach the same fraction of the connected network with 5% failure as with zero failure (Fig. 1c). Targeted attacks have a much greater impact on the exponential network, eventually causing cascades sizes to drop to zero; however, exponential networks do not have a critical transition for complex contagions, and are relatively robust even after losing the 100 most connected nodes (1%). This tolerance of exponential networks suggests that in social networks the reinforcement of desired norms can be sustained despite continual network attrition due to death or mobility.

1. Albert, R., Jeong, H. and Barabasi, A.L. *Nature* 406, 379-381 (2000).

2.. Holme, P., & Kim, B. J. *Phys. Rev. E.* 65, 056109 (2002).

3. Barabasi, A.L, & Albert, R. *Science* 286, 509-511 (1999).

4. Redner, S. *Euro Phys. J. B*. 4, 131-134 (1998).

5. Centola, D., Eguiluz, V., Macy, M. *Physica A* (in press).

6. McAdam, D. & Paulsen, R. *Am. J. of Soc.* 99, 640-667 (1993).

7. Newman, M.E.J. & Park, J. *Phys. Rev. E.* 68, 036122 (2003).

8. Newman, M.E.J., *Proc. Nat. Acad. Sci. U.S.A.* 98, 404-409 (2001).

9. Newman, M.E.J., *SIAM Review* 45, 167-256 (2003).

Figure 1. Spread of simple and complex contagions on scale-free and exponential networks. a, Minimally complex contagions require that each node have 2 (but not more than 100%) of their neighbours activated in order to become activated. This can cause "bottlenecks" where a contagion (blue nodes) cannot spread to reach every node in a connected network. b, Average size of cascades of complex contagions as fraction $f$ of nodes are removed from a clustered scale-free network by random failure (dotted line) or targeted attack (solid line), and the connectedness of the network (solid line with circles) as the size of cascades of simple contagions. c, Average size of cascades on an exponential network for complex contagion with random failure (dotted line) and targeted attack (solid line), and for simple contagion (solid line with circles).

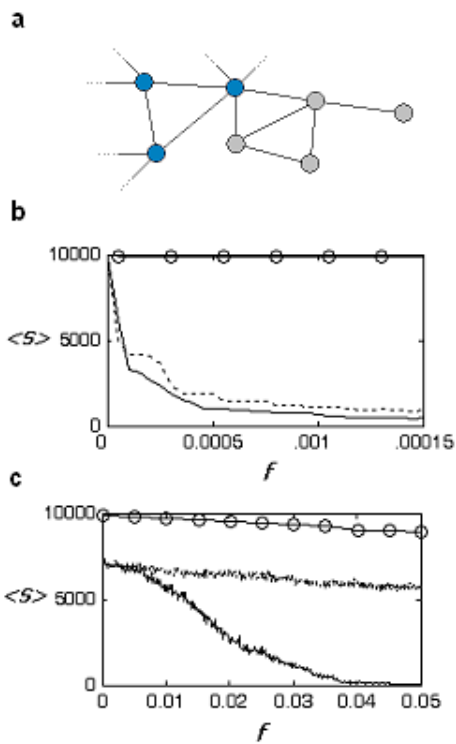